\documentclass[nofootinbib,twocolumn]{revtex4-1}
\usepackage{graphicx}   
\usepackage{amsmath}    
\usepackage{amssymb}    
\usepackage{mathtools}
\usepackage{upgreek}
\usepackage{bm}         
\usepackage{fancyvrb}   
\usepackage{color}      
\usepackage{subfigure}  
\usepackage{hyperref}   
\usepackage{titlesec}
\usepackage{float}
\def\Re{\mathrm{Re}}
\def\cO{\mathcal{O}}
\def\d{\mathrm{d}}
\newcommand\pd{\partial}

\begin{document}

\title{The image of a point charge in an infinite conducting cylinder}

\author{Matt Majic (mattmajic@gmail.com)}
\affiliation{The MacDiarmid Institute for Advanced Materials and Nanotechnology,
School of Chemical and Physical Sciences, \\
Victoria University of Wellington,
PO Box 600, Wellington 6140, New Zealand}

\begin{abstract}
The electrostatics problem of a point charge next to a conducting plane is best solved by placing an image charge placed on the opposite side. For a charge between two parallel planes this can be solved with image charges outside the planes at evenly spaced intervals moving out to infinity. What is the corresponding image of a point charge is when placed on the axis of a cylinder?. The potential of a point charge in a cylinder is well known and may expressed in many forms involving integrals or series of Bessel functions, but none of which elude to an image. In fact the image consists of infinitely many rings on a disk with some complicated surface charge distribution. We attempt to describe the image as accurately as possible, and in doing so find simple accurate approximations for the potential, and derive an expression for the image charge density.
\end{abstract}
\maketitle

The method of images is an approach to a physical problem comprised of a source potential and surface boundary. It may provide a simple expression for the reflected potential when other methods involving series, integrals or surface charges on the conductor are not practical. Some electrostatic problems may be solved by placing image charges successively where each image generates another, and eventually this process converges to satisfy the boundary conditions.
The most basic example is that of a point charge next to an infinite conducting plane, where the image is a negative point charge located opposite the plane. This is just like the image in a mirror, but with the additional consideration of sign. The analogy arises because these two cases are the long- and short-wavelength limits of light scattering governed by the Helmholtz equation, where the same image solution applies regardless of wavelength. In very few cases are scattering problems for the Helmholtz equation solvable with images, but for electrostatics - Laplace's equation - there are many examples. For multiple planes of any orientation an image solution can be constructed knowing the solution for a single plane. The image of a point charge in a conducting sphere is the Kelvin point charge. For a spheroid, Ref. \cite{dassios2012neumann} used an (unreduced) image solution involving a curved line of charge enclosed in a charged spheroidal surface, but the reduced form of the image is unknown, even in the case of axial incidence \cite{lindell2001electrostatic}.
For a circular disk an image can be constructed by considering a second copy of the entire space attached at the disk, where the potential of the point charge in this double space is modified - the image is then another one of these modified point charges located opposite the disk but in the second space \cite{davis1975solution}. A similar solution applies for a half-plane. The potential of two charged conducting spheres can be solved using an infinite series of image point charges \cite{smythe1988static}. A more complicated image approach can be applied to materials with finite permittivity, for example the image of a point charge in a dielectric sphere is a line extending radially from the center. But the exact image of a point charge near a cylindrical surface has not been found, although an approximate numerical approach using point images has been implemented \cite{xu2011image}. The simplest case is for the charge placed inside the cylinder exactly on the axis - in fact here there are no variable parameters other than scaling factors. Even in this case we find that the image
\onecolumngrid
\begin{center}
	\begin{figure}[b]
		\includegraphics[scale=0.31]{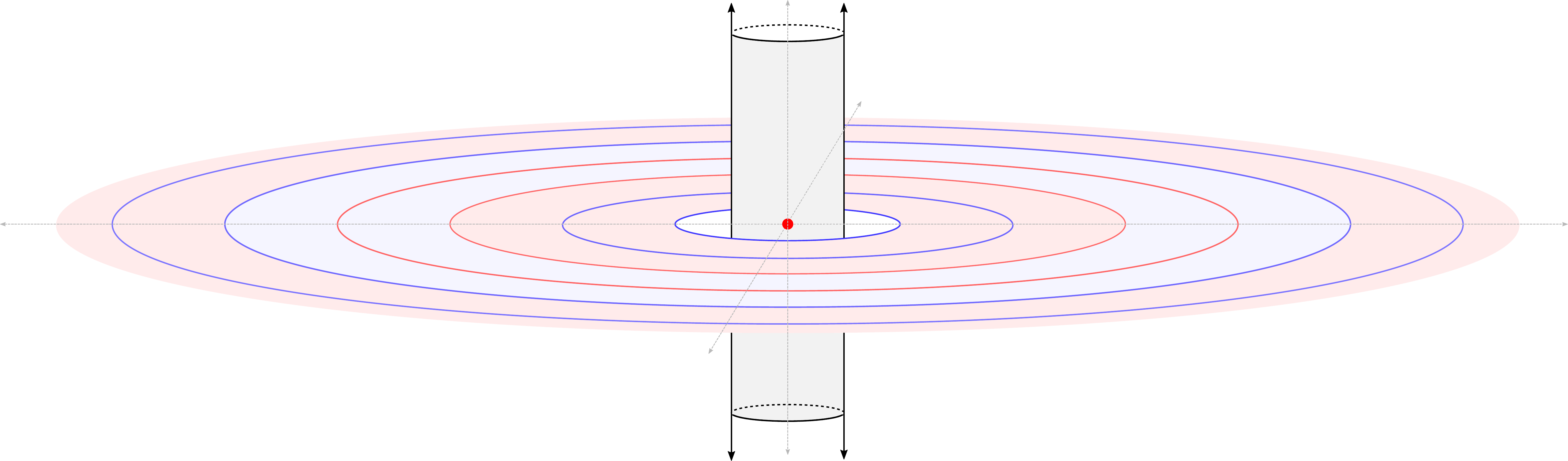}\\
		\caption{Schematic of the problem with of the point charge, cylinder and a rough illustration of the image disk, with positive charge in red and negative in blue. The image extends to infinity and has a near-alternating pattern with evenly spaced singular rings.}
		\label{schematic}
	\end{figure}
\begin{figure*}
	\includegraphics[scale=0.62]{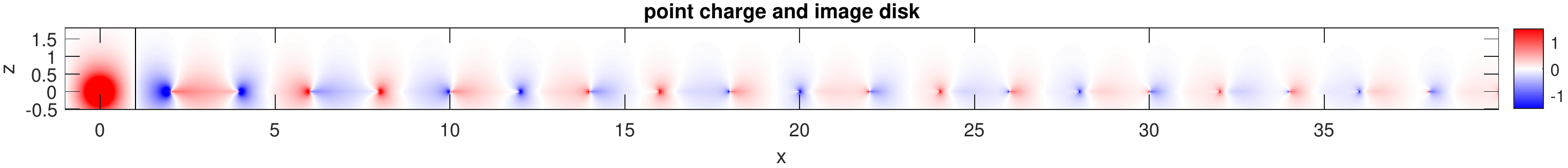}
	\caption{The total potential $V$ showing the point charge and image disk, calculated with the series \eqref{besselsum} with 500 terms, more than necessary to converge to visual accuracy. The black line at $x=1$ is the cylinder.}
	\label{imagedisk}
\end{figure*}
\end{center}
\twocolumngrid
\noindent is surprisingly complex, consisting of an infinite number of rings on a disk with a complicated surface charge distribution. In particular we manage to prove that the image disk is singular at evenly spaced concentric rings, and find some simple analytic approximate images which provide accurate approximations to the potential. Finally in section \ref{density} we derive a series expression for the image charge density.

\section{Problem and series solution}

Consider a point charge on the axis of a conducting infinite cylindrical tube radius 1, where the total electrostatic potential inside is $V$.
The complete analytic continuation of $V$ is derived in \cite{dougall1900determination}, \cite{bouwkamp1947electrostatic}, expressed as a series of Bessel functions:
\begin{align}
V=2\sum_{n=1}^\infty\frac{J_0(k_n\rho)e^{-k_n| z|}}{k_nJ_1(k_n)^2} \label{besselsum}
\end{align}
where $J_0$ and $J_1$ are Bessel functions of the first kind and $k_n$ is the $n^\text{th}$ zero of $J_0$. This series accounts for both the point charge and the potential reflected by the cylinder. Physically the solution is only needed inside the cylinder and one simply ignores $V$ outside, but mathematically \eqref{besselsum} can be evaluated outside to reveal the virtual ``image" of the point charge. 
A schematic of the cylinder and the general structure of the image is shown in figure \ref{schematic}, and $V$ is plotted more accurately in figure \ref{imagedisk} on a cut of the $xz$ plane. The plot required many terms (500) to evaluate due to the series being slowly convergent for small $z$ and conditionally convergent for $z=0$, meaning the sum of absolute values diverges. It appears that the image lies on the disk $\rho\geq2,~z=0$, and also diverges on rings at $\rho=2, 4, 6, 8$ ... . The singular rings diverge to both positive and negative values on either side, in a different way for each ring. The types of singularity appear to roughly repeat every 4 rings - negative on the inside, negative on the outside, positive on the inside, positive on the outside, repeat. And the charge appears to decay gradually as $\rho\rightarrow\infty$.



\section{Comparison to the two planes problem}
\begin{figure}[h!]
\includegraphics[scale=0.3]{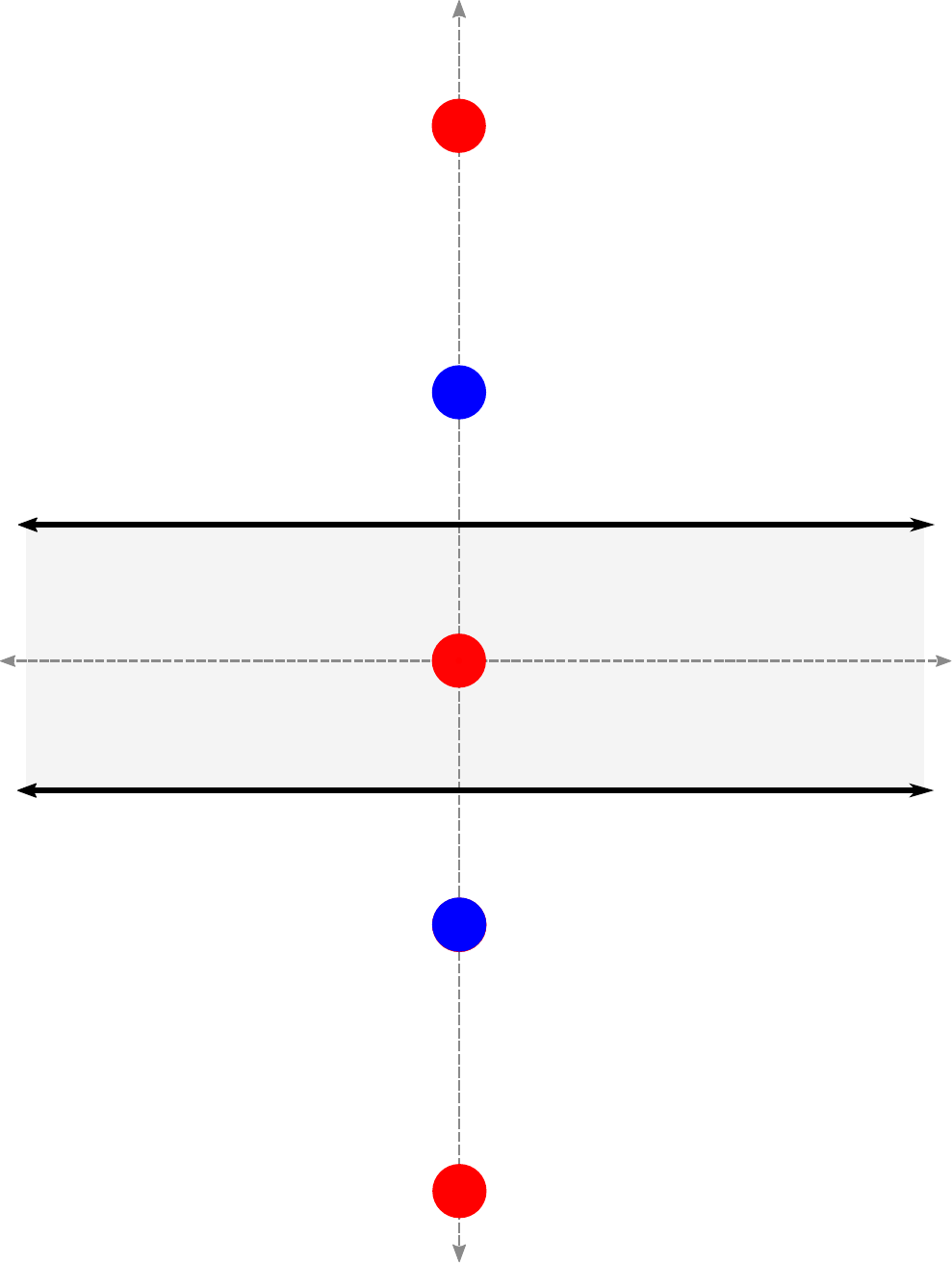}
\caption{Images of a point charge between two planes.} \label{TwoPLanesImages}
\end{figure}
The image disk shares a similarity with the image solution for the problem of a point charge at the center of two parallel conducting planes, lying at $z=\pm 1$. For this problem the image is made of point charges located at $z=\pm2$ with charge -1, $z=\pm 4$ with charge +1, $z=\pm 6$ with charge -1, out to infinity:

\begin{align}
V_\text{planes}=\sum_{k=-\infty}^\infty \frac{(-)^k}{\sqrt{\rho^2+(z-2k)^2}}. \label{twoplanesimages}
\end{align}
These images are shown schematically in figure \ref{TwoPLanesImages}. 
An alternate approach using cylindrical harmonics was employed in \cite{dougall1900determination} to obtain
\begin{align}
V_\text{planes}=\frac{\pi}{2}\sum_{n=1,\text{ odd}}^\infty \cos\left(\frac{n\pi z}{2}\right) K_0\left(\frac{n\pi \rho}{2}\right). \label{twoplanesbessel}
\end{align}
where $K_0$ is the modified Bessel function of the second kind.
This is of a similar form to \eqref{besselsum},  but with the function arguments increasing in integer intervals instead of the zeros $k_n$. This is due to the fact that the zeros of the cosine function lie at even intervals. Also, in \eqref{twoplanesbessel} the sum involves cylindrical harmonics of imaginary separation parameter, which diverge on the $z$-axis.

Later we will show that the first order approximation to the cylinder's image rings are actually point charges located at $z= \pm2i,\pm4i,\pm6i...$. So the image disk shares this regular pattern of singularities, but the ring charges alternate sign/orientation every 4 rings, not every 2, and there is surface charge between the rings with a somewhat non-uniform pattern.

In the two planes problem, the image charges can be explained intuitively by considering them two at a time moving outwards. Each image charge induces another image of opposite sign reflected about the plane furthest from it. We can attempt to apply this explanation for the cylinder - consider one point on an image ring and the opposite side of the cylinder, and imagine a tangent plane to that edge - then we see that at least this infinitesimal edge will induce an image of that point on the next ring, an extra distance of 2 away, just as done for the plane. But this explanation only applies to an infinitesimal part of the cylinder.


\section{Approximation of the series (\ref{besselsum})}

\begin{figure}[h!]
	\includegraphics[scale=.31,trim={0 0 0 3cm}, clip]{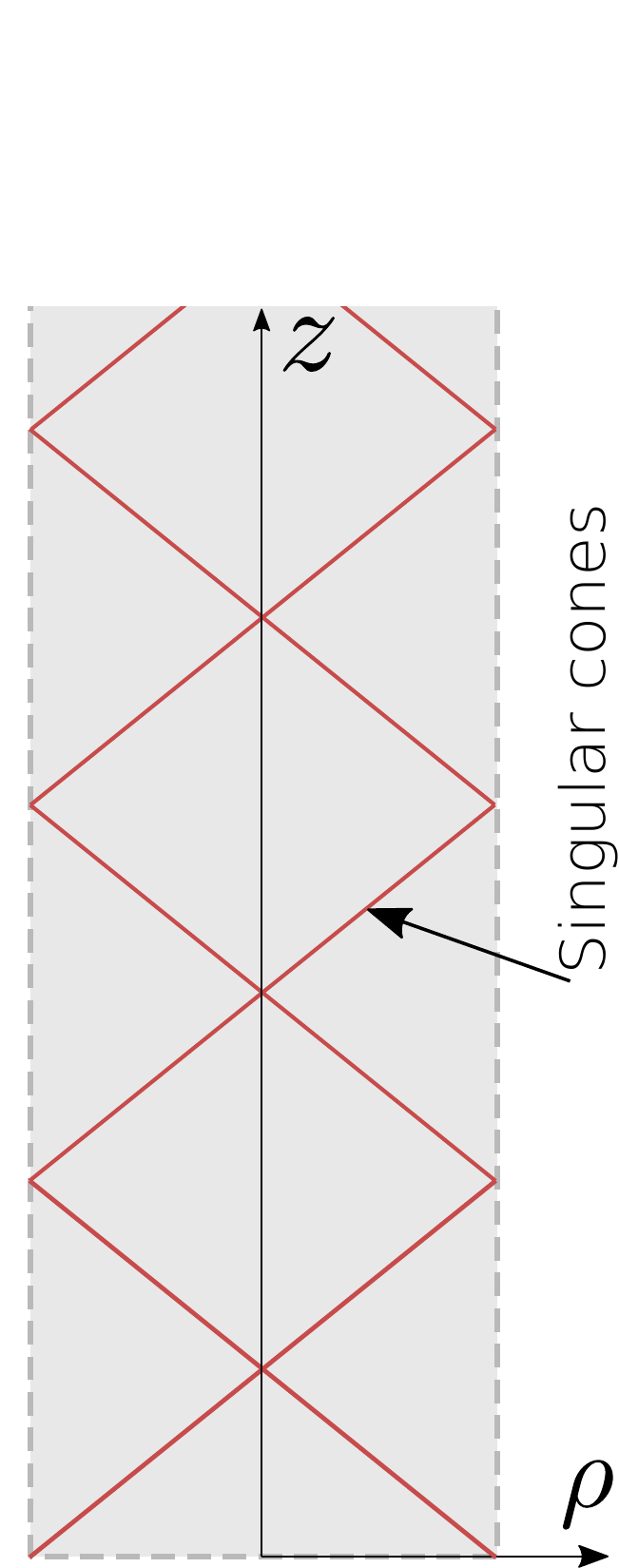}
	\caption{Locations of the singularities for a jet flow confined to a semi infinite cylinder.}
	\label{jet singularities}
\end{figure}

A similar series to \eqref{besselsum} was encountered in Ref. \cite{dillmann1995method}, in computing the velocity field of an axisymmetric jet flow confined to a semi-infinite cylinder. and was dealt with using Kummer's method of series acceleration - subtracting a similar analytic series whose terms behave the same way as the summation index goes to infinity, so that the remaining series converges faster. So we will briefly cover this paper and its technique.
They considered a jet confined to $0\leq \rho\leq 1$ and $z>0$, and wished to compute for example the axial component of the fluid velocity, which is proportional to the series
\begin{align}
	S_z=2\sum_{n=1}^{\infty}\frac{J_0(k_n \rho)\cos(k_n  z)}{k_n J_1(k_n)}. \label{jet series}
\end{align}
The terms decrease slowly as $1/n$ so convergence is conditional and nonuniform across both $\rho$ and $z$, \textit{everywhere} inside the jet - which necessitates a series acceleration technique for practical computation. In fact, the singularities of \eqref{jet series} lie in real space - on evenly spaced cone shaped surfaces running along the jet as shown in figure \ref{jet singularities}.  This series differs to our \eqref{besselsum} by one factor of $J_1(k_n)$ and $e^{-k_n z}\rightarrow\cos(k_n z)$, and both series have similar singularities, but in their case they were interested in singularities at integer spaced values of $z$, not $\rho$. 

The approach of \cite{dillmann1995method} used the following asymptotic formulas for $n\rightarrow\infty$:
\begin{align}
k_n =& \bigg(n- \frac{1}{4}\bigg)\pi +\frac{1}{8\pi n} + \cO(n^{-2}),\\
J_1(k_n)=& \frac{(-)^{n+1}}{\pi}\sqrt{\frac{2}{n\!-\!1/4}}  +\cO(n^{-7/2}), \\
J_0(k_n\rho)=& \frac{(-)^n}{\pi}\sqrt{\frac{2}{(n\!-\!1/4)\rho}}\bigg\{\sin[(n\!-\!1/4)\pi(\rho-1)] \nonumber\\
& - \frac{\rho -1/\rho}{8(n\!-\!1/4)\pi}\cos[((n-1/4)\pi(\rho-1)] \bigg\} \nonumber\\
&\hspace{4cm} + \cO(n^{-5/2}).
\end{align}
The series $S_z$ could then be approximated by a simpler trigonometric series related to the Lerch transcendent which could be evaluated analytically. The remainder series converges much faster and is bounded. 

We now follow the same approach for our problem to extract the singularities of $V$. To first order in $n$ the series coefficients behave as
\begin{align}
2&\frac{J_0(k_n \rho)}{k_n J_1(k_n)^2}= (-)^n \sqrt{\frac{2}{(n\!-\!1/4)\rho}}\sin[(n-1/4)\pi(\rho-1)]\nonumber\\
&\hspace{5cm}+ \mathcal{O}(n^{-3/2}). \label{asymptotic}
\end{align}
The divergent part of $V$ is then entirely contained within the order $n^{-1/2}$ part, since the remaining series converges as $n^{-3/2}$ which is absolutely convergent (although it still contains discontinuities in the derivative at $\rho=2,4,6...$ which can be dealt with by considering the next order in the asymptotic expansion). So the poles of $V$ coincide with the poles of 
\begin{align}
	V'=&\sum\limits_{n=1}^{\infty} (-)^n \sqrt{\frac{2}{n\rho}}\sin[(n-1/4)\pi(\rho-1)]e^{-(n-1/4)\pi z} \label{V'}
	\end{align}
for $z>0$. While $V'$ does not solve Laplace's equation, it can still be used to determine properties of $V$. It is clear that $\sqrt{\rho}V'$ is periodic in $\rho$ with period 8, and diverges for $\rho=2,4,6...$ - these are the stationary points where $\sqrt{2}(-)^n\sin[(\rho-1)(n-1/4)\pi]=1$, which results in the divergent series $\sum_{n\geq 1}n^{-1/2}$.  This proves that the singularities of $V$ also lie on $\rho=2,4,6...$ and follow a repeating pattern every 4 singularities, and decrease in magnitude moving out from the origin as $1/\sqrt{\rho}$ which agrees with the plot in figure \ref{imagedisk}. To make further deductions, we will rewrite $V'$ using an addition formula for the sine function and the polylogarithm of index $\frac{1}{2}$,
\begin{align}
	L_\frac{1}{2}(e^{\pi\mu})=\sum_{n=1}^\infty\frac{e^{n\pi\mu}}{\sqrt{n}},
\end{align}
to obtain
\begin{align}
 V' =&\frac{i e^{(i\rho-i-z)\frac{\pi}{4}}}{\sqrt{2\rho}}\big[ e^{i(\rho-1)\frac{\pi}{2}}L_\frac{1}{2}( e^{(z-i\rho)\pi}) - L_\frac{1}{2}( e^{(z+i\rho)\pi}) \big], \label{polylog}
\end{align}
 which is real valued. This form shows that away from the singularities, $V'$ is continuous due to the convergence of the polylogarithm expressed as a Bose-Einstein integral.
Also from looking at the limit as $\rho\rightarrow2k$ in Eq. \eqref{polylog}, we can determine simple analytic approximations for the $k^{th}$ singular ring. This limit may be found from the residue series expression for the polylogarithm \cite{wood1992computation}:
\begin{figure*}
	\includegraphics[scale=.683]{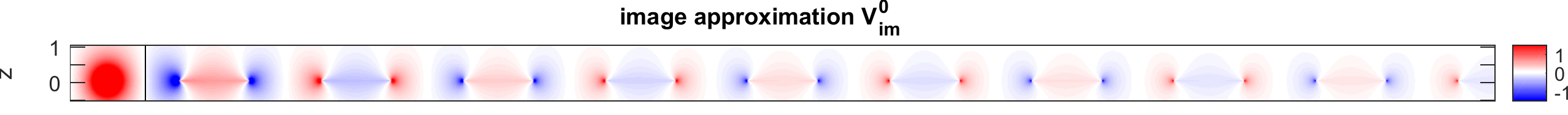}~$^\text{(a)}$\\
	\includegraphics[scale=.683]{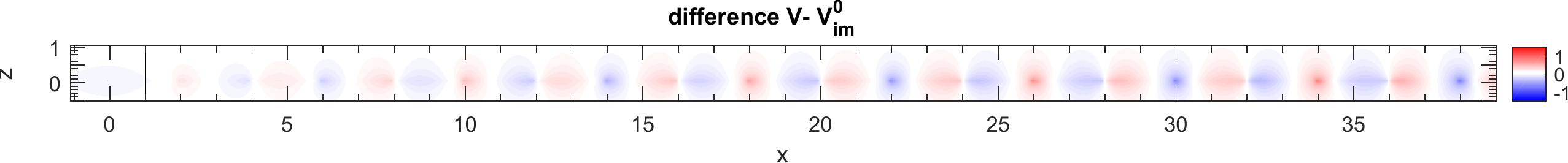}$^\text{(b)}$
	\caption{(a): the image approximation potential $V_\text{im}^0$ computed up to $k=160$ images showing a similar structure to that for $V$. (b): the difference $V-V_\text{im}^0$. The black line at $x=1$ is the cylinder. The same color scale is used in both plots.}
	\label{Vimages plot}
\end{figure*}
\begin{align}
L_\frac{1}{2}(e^{\pi\mu})=\sum_{q=-\infty}^\infty \frac{1}{\sqrt{2iq-\mu}}. \label{Bose}
\end{align}
For the limit approaching ring $k$, the term $q=\pm k$ dominates, leaving
\begin{align}
\lim_{\substack{\rho\rightarrow2k\\z\rightarrow0}} V' = \frac{i^k}{2\sqrt{k}}\left(\frac{1}{\sqrt{\rho+iz-2k}} + \frac{(-)^k}{\sqrt{\rho-iz-2k}}\right). \label{V'limits}
\end{align}
In section \ref{sec first order} it is shown that a combination of 2 point charges located at $z=\pm2i$ is an accurate approximation to the first ring. In fact, on the $z=0$ plane these point charges share the same limit as  \eqref{V'limits} for $k=1$. From this we can assume that similar point charges can be used to match the singularities for the higher rings. Explicit expressions for these image point charges are
\begin{align}
 V_{\text{im,}k}^0 = &\frac{i^k }{\sqrt{\rho^2+(z-2ik)^2}} + \frac{(-i)^k }{\sqrt{\rho^2+(z+2ik)^2}}.\label{Vimk}
\end{align}
It is straightforward to show that $V_{\text{im,}k}^0$ matches the limits in \eqref{V'limits} for any $k$.  
Then summing these together gives an approximation for the potential: 
\begin{align}
V\approx V_\text{im}^0=&\frac{1}{r} + \sum_{k=1}^\infty V_{\text{im,}k}^0, \label{Vim}
\end{align}
$V_{\text{im,}k}^0 $ are real valued and singular on the rings but also posses discontinuities on the $z=0$ plane. In particular, for $k$ even(odd), $V_{\text{im,}k}^0$ is discontinuous on the inner(outer) disk $\rho<2k~(\rho>2k)$. So $V_\text{im}^0$ is discontinuous inside the cylinder and not a practical approximation to $V$. In figure \ref{Vimages plot}, $V_\text{im}^0$ is plotted along with the difference relative to $V$. While the difference is not negligible, it is finite everywhere.  Figure \ref{Vimages plot} (b) shows increasing error for the outer rings, but tests out to $\rho=1000$ indicate that the error does eventually decrease, consistent with the approximation \eqref{V'} becoming more accurate as $\rho\rightarrow\infty$.

By comparing the series for $V_\text{im}^0$ to the image series solution for the two planes problem \eqref{twoplanesimages}, and noting that this is also equal to the Bessel series \eqref{twoplanesbessel}, we can derive with some algebra the following Bessel series for $V_\text{im}^0$:
\begin{align}
	V_\text{im}^0=\pi\sum_{n=1}^\infty e^{-(n-1/4)\pi |z|} J_0((n-1/4)\pi\rho) \label{Vim0bessel}
\end{align}
which more closely resembles the series \eqref{besselsum}, with nonuniform convergence as $n^{-1/2}$ for $z=0$. This series has the same asymtotic limit as $n\rightarrow\infty$ as the series for both $V$ and $V'$.

\section{Integral solution}
Towards finding a more practical approximation to $V$, we start with the integral solution, which involves splitting the potential into $V=V_e+V_r$ where $V_e$ is the excitation of the point charge and $V_r$ is reflected by the cylinder. $V_e$ can be expressed as an integral of cylindrical harmonics:
\begin{align}
	V_e=\frac{1}{r}=\frac{2}{\pi}\int_0^\infty K_0(t\rho)\cos(t z)\d t, \label{Ve}
\end{align}
where $I_0$ and $K_0$ are the modified Bessel functions of the first and second kinds. $V_r$ is constructed to fit the boundary condition $V=0$ at $\rho=1$ \cite{bouwkamp1947electrostatic}:
\begin{align}
	V_r=-\frac{2}{\pi}\int_0^\infty \frac{K_0(t)}{I_0(t)}I_0(t\rho)\cos(t z)\d t . \label{Vr}
\end{align}
The integrand is finite except at $t\rightarrow0,\infty,$ so we can determine the physical domain where this integral converges by analysis of the limiting behavior of the integrand. The Bessel functions behave as
\begin{align}
I_0(t\rightarrow0)&\rightarrow 1\\
K_0(t\rightarrow0)&\rightarrow \log\frac{1}{t}\\
I_0(t\rightarrow\infty)&\rightarrow \frac{e^t}{\sqrt{2\pi t}}\bigg(1+\frac{1}{8t}\bigg)\\
K_0(t\rightarrow\infty)&\rightarrow  e^{-t}\sqrt{\frac{\pi}{2 t}}\bigg(1-\frac{1}{8t}\bigg).
\end{align}

Then as $t\rightarrow\infty$ the integrand behaves as $e^{(\rho-2)t}/\sqrt{t}$ and converges for $\rho<2$, independent of $z$, inside a cylindrical boundary of twice the radius of the physical cylinder. This is consistent with the image having its innermost ring at $\rho=2$.


\section{Spherical series solution}
The integral solution may be transformed into a series of solid spherical harmonics, which will then be useful to determine approximations to $V_r$. The coefficients can be obtained via the following expansion relating cylindrical and spherical harmonics:
\begin{align}
	I_0(t\rho)\cos(tz)=\sum_{n=0:2}^\infty\frac{(itr)^n}{n!}P_n(\cos\theta), \label{cylindrical vs spherical}
\end{align}
where the notation $n=0:2$ means the summation only covers $n$ in steps of 2. Substituting \eqref{cylindrical vs spherical} into the integral \eqref{Vr} and rearranging gives
\begin{align}
V_r=-\sum_{n=0:2}^\infty h_n \left(\frac{ir}{2}\right)^n P_n(\cos\theta), \label{Vr spherical}
\end{align}
where 
\begin{align}
		h_n&=\frac{2^{n+1}}{\pi n!}\int_0^\infty \frac{K_0(t)}{I_0(t)}t^n\d t. \label{hn2}
\end{align}

Note that in \cite{bouwkamp1947electrostatic,greenwood1968simple}, a different expression for $h_n$ is derived:
\begin{align}
h_n&=\frac{1}{n+1}\frac{2^{n+1}}{\pi n!}\int_0^\infty \frac{t^n}{I_0(t)^2}\d t, \label{hn1}
\end{align}
which is equivalent to \eqref{hn1} through partial integration and applying the Wronskian $I_0(t)K_0'(t)-I_0'(t)K_0(t)=-1/t$.



\section{A first order approximation} \label{sec first order}
\begin{figure}
	  \includegraphics[scale=0.65]{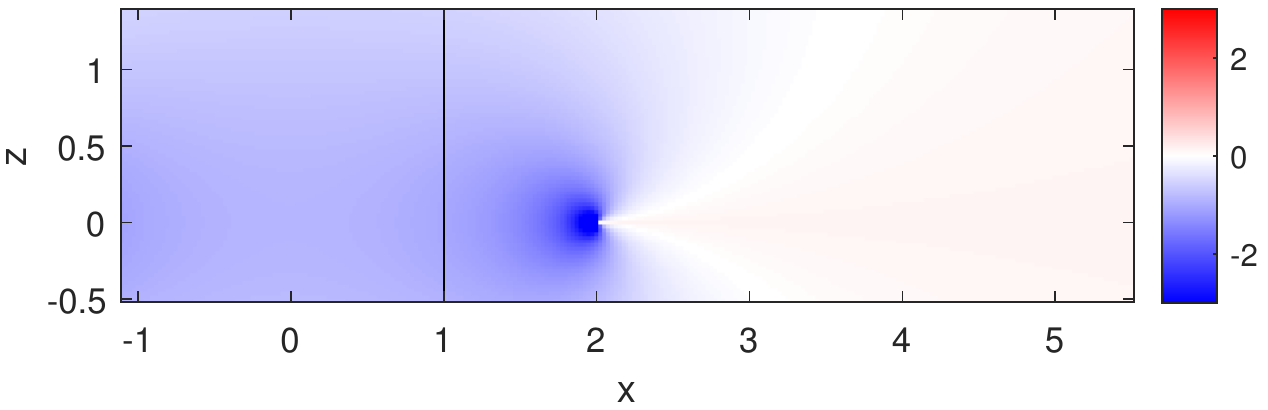}~$^\text{(a)}$
	  \includegraphics[scale=0.65]{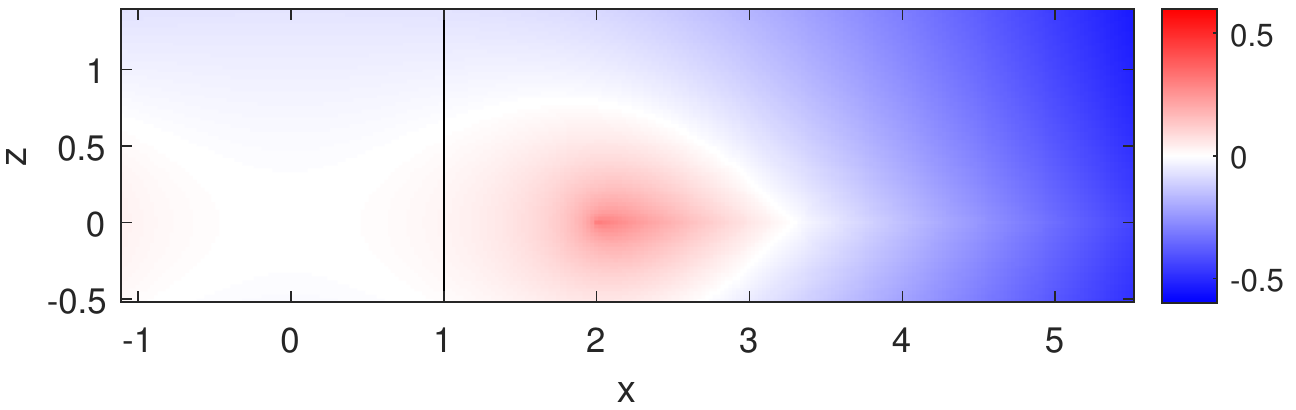}$\!\!\! ^\text{(b)}$
	  \includegraphics[scale=0.65]{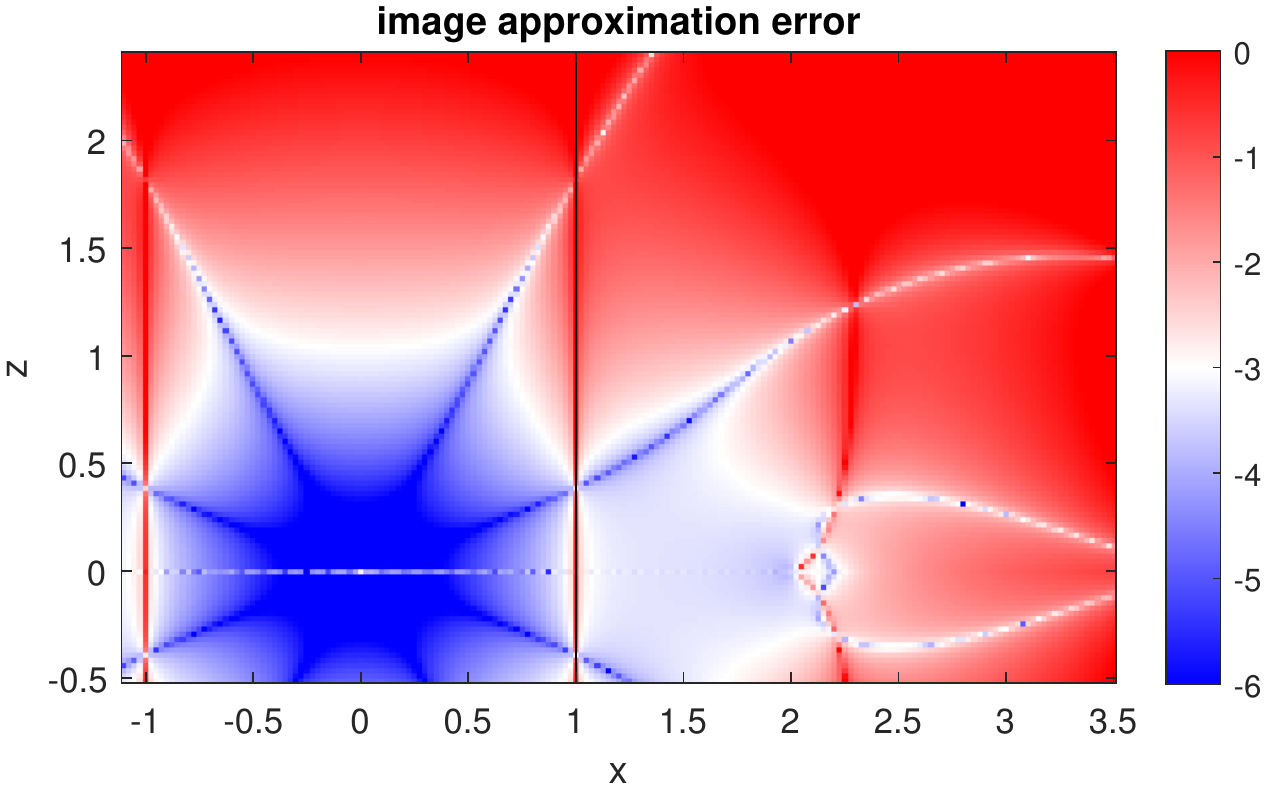}$\!\!\! ^\text{(c)}$
	\caption{(a): approximation $V^{(0)}$, (b): correction $V^{(1)}$, (c): $\log_{10}$ of the relative error $(V^{(0)}+V^{(1)}-V)/V$.}
	\label{imagedisks}
\end{figure}

\begin{figure*}
	\includegraphics[scale=0.66]{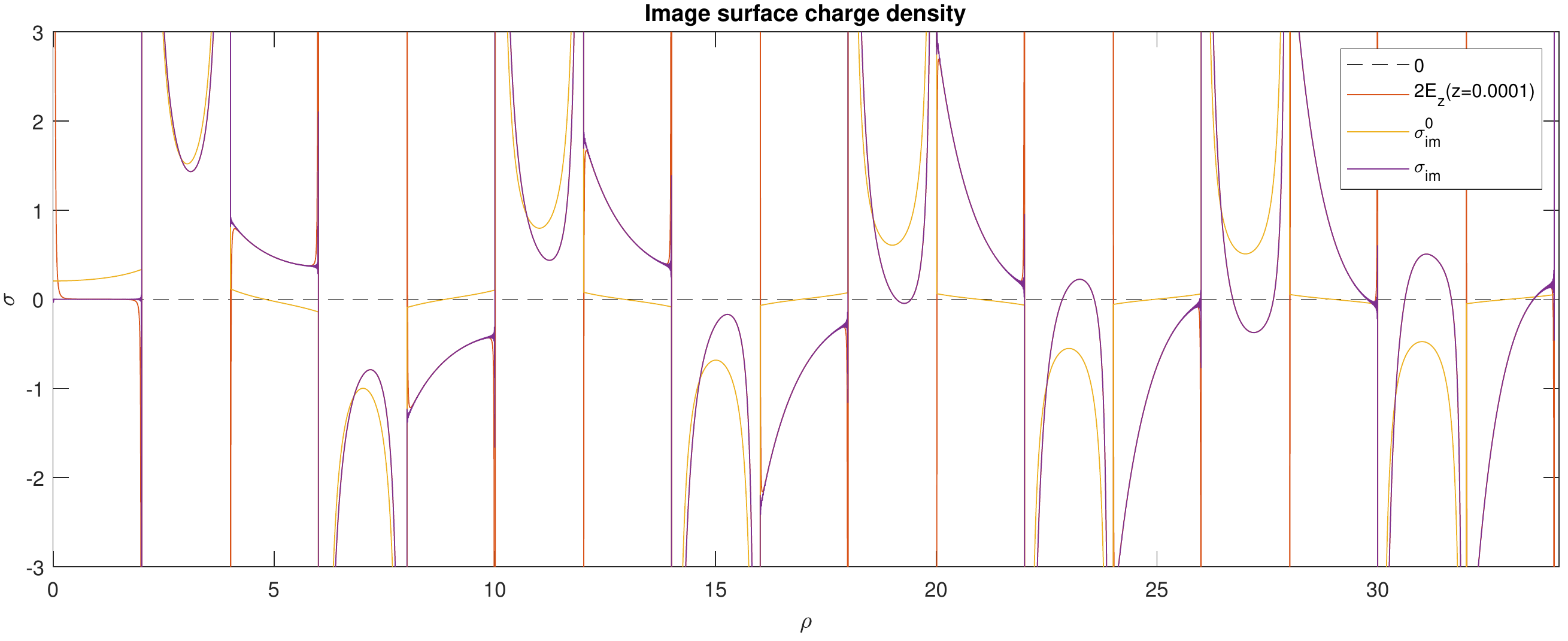}
	\caption{The surface charge density $\sigma_\text{torus}$ compared to the approximation $\sigma_\text{torus}^0$ and the electric field $E_z$ evaluated at $z=0.0001$. $E_z$ and $\sigma_\text{torus}$ are computed with 20,000 terms in the Bessel series \eqref{E_z torus} and \eqref{sig torus}, while $\sigma_\text{torus}^0$ is computed with 80 terms in the series \eqref{sig0 torus}, enough to converge to visible accuracy. } \label{surface charge}
\end{figure*}

From the spherical series solution \eqref{Vr spherical} we can find a simple approximation that happens to match the image well. The integrand of \eqref{hn2} may be expanded for $t\rightarrow\infty$ as:
\begin{align}
	\frac{K_0(t)}{I_0(t)}t^n\rightarrow \pi\bigg(1-\frac{1}{4t}+\cO(t^{-2})\bigg)e^{-2t}t^n,
\end{align}
which also is fairly accurate for small $t$. Because of the factor $t^n$, as $n\rightarrow\infty$ the contributions in the integral only come from large $t$. This approximation leads to $h_{n\rightarrow\infty}\rightarrow 1-\frac{1}{2n}$.

The zeroth order image has $h_n\approx 1$. The first two terms for $n=0,2$ are however not well represented by this approximation so should be subtracted and the exact terms with $h_0\approx0.8706901$ $h_2\approx0.8236450$ and added on explicitly:
\begin{align}
	V^{(0)}=& - \sum_{n=4:2}^{\infty}\bigg(\frac{r}{2i}\bigg)^{n}P_{n}(\cos\theta) -h_0-h_2r^2P_2(\cos\theta) \nonumber\\
	=& 
	 \frac{i}{\sqrt{\rho^2+(z+2i)^2}} - \frac{i}{\sqrt{\rho^2+(z-2i)^2}} \nonumber\\
	&+1 - h_0 - \bigg(\frac{1}{4}-h_2\bigg)r^2P_2(\cos\theta), \label{Vim0} 
\end{align}
which is two point charges located at $z=\pm2i$ that appear in $V_\text{im1}^0$, plus smooth terms. 
$V^{(0)}$ is real and singular at $\rho=2,~z=0$ and is discontinuous across $\rho\geq2,~z=0$, as seen in figure \ref{imagedisks} (a), with surface charge $4/(\rho^2-4)^{3/2}$. Just like $V_\text{im1}^0$, $V^{(0)}-V$ is finite at the innermost ring. This approximation was also derived in \cite{bouwkamp1947electrostatic} but with the wrong prefactor and without correction of the $n=0,2$ terms.

For the first order correction again we take off the $n=2$ term:
\begin{align}
		V^{(1)}=&  \sum_{n=4:2}^{\infty}\frac{1}{2n}\bigg(\frac{r}{2i}\bigg)^{n}P_{n}(\cos\theta) \nonumber\\
		=& \frac{1}{2}\text{Re}\bigg\{\log\frac{2}{2-iz+\sqrt{-\rho^2-(z+2)^2}}\bigg\} \nonumber\\
		&+\frac{1}{16}r^2P_2(\cos\theta).
\end{align}
This is also discontinuous across $\rho\geq2,~z=0$. 
The logarithm can be derived from a coordinate transformation applied to the expansion of a similar function in \cite{ranvcic2006point}. $V^{(1)}$ has a positive surface charge density of $1/\sqrt{\rho^2-4}$ on the disk $\rho\geq2,z=0$, but also consists of negative charge at $r\rightarrow\infty$ so that even on the disk, the potential can be negative, as seen in figure \ref{imagedisks} (b). The potential is finite everywhere even at $\rho=2,z=0$, but diverges logarithmically as $r\rightarrow\infty$.

The approximation $V^{(0)}+V^{(1)}$ is very accurate for $|z|\lesssim1.5$, as shown in figure \ref{imagedisks} (c). Numerical evidence suggests that difference $V_r-(V^{(0)}+V^{(1)})$ and its low order derivatives at least are continuous across the innermost image ring.

This seems to be the extent of these image approximations since this expansion of $h_n$ about $n\rightarrow\infty$ appears to not match well for small $n$. Already in our approximation we have had to correct for $n=0,2$, and more corrections are likely needed for higher orders if continuing in this fashion. 
And unfortunately adding similar terms for the next ring at $\rho=4$ only degrades the approximation - as noted before $V_\text{im,2}^0$ is discontinuous across the middle of the cylinder.

\section{Image surface charge density}\label{density}

To complete the description of the image disk, we must determine its charge distribution.
The surface charge is equivalent to the discontinuity in the electric field component $E_z=-\pd_z V$ across $z=0$. Since the series for $E_z$ at $z=0$ actually diverges - the terms increase as $\sqrt{n}$ - we must instead consider the limit:
\begin{align}
\sigma_\text{im} = 2\lim_{z\rightarrow0}E_z =-\lim_{z\rightarrow0} 4\sum_{n=1}^\infty\frac{J_0(k_n\rho)e^{-k_n z}}{J_1(k_n)^2}. \label{E_z}
\end{align}

In order to calculate this limit, we can modify the series using Kummer acceleration - subtract off an analytically known term that has a similar series expansion, so that the difference of the two series coefficients decreases faster as $n\rightarrow\infty$. 
We will in fact subtract off the surface charge density of the approximation $V_\text{im}^0$. This can be obtained from either \eqref{Vim}:
\begin{align}
\sigma_\text{im}^0=8\sum_{k=1}^\infty \Re\bigg\{\frac{i^kk}{(4k^2-\rho^2)^{3/2}} \bigg\}\label{sig0}
\end{align}
(The direct substitution of $z=0$ in $\pd_z$\eqref{Vimk} leads to a different branch cut, while \eqref{sig0} provides the correct limit - thanks to Carlo Beenakker on MathOverflow for providing \eqref{sig0}), or from the Bessel series \eqref{Vim0bessel}:
\begin{align}
\sigma_\text{im}^0 = -\lim_{z\rightarrow0} 2\pi^2\sum_{n=1}^\infty J_0((n-1/4)\pi\rho)e^{-(n-1/4)\pi z} \label{sig0 divergent}
\end{align}

\begin{figure*}
	\includegraphics[scale=0.66]{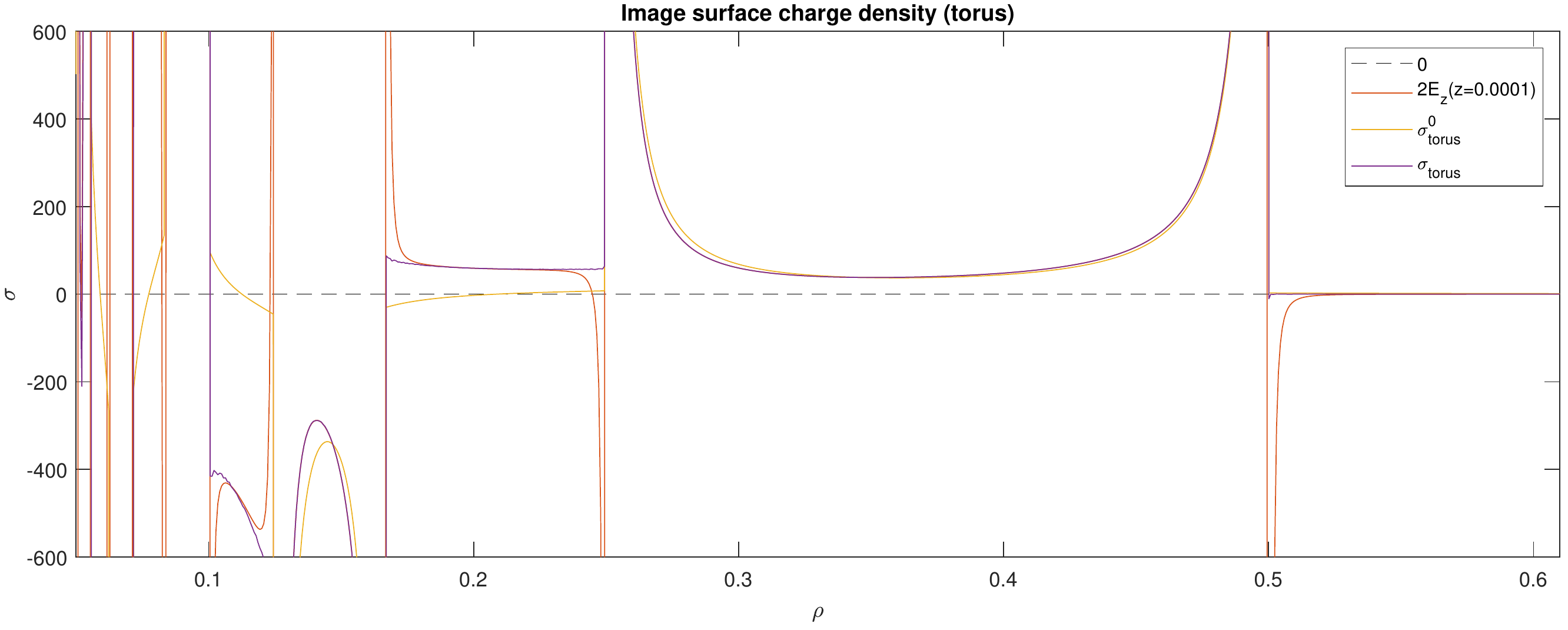}
	\caption{The surface charge density $\sigma_\text{torus}$ compared to the approximation $\sigma_\text{torus}^0$ and the electric field $E_z$ evaluated at $z=0.0001$. $E_z$ and $\sigma_\text{torus}$ are computed with 50,000 terms in the Bessel series \eqref{E_z torus} and \eqref{sig torus}, while $\sigma_\text{torus}^0$ is computed with 500 terms in the series \eqref{sig0 torus}, enough to converge to visible accuracy. The plot is truncated at $x=0.05$ since the information for $x<0.05$ is illegible.}  \label{surface charge torus}
\end{figure*}

 which is again divergent for $z=0$, with terms increasing as $\sqrt{n}$. In fact, as $n\rightarrow\infty$, the series coefficients in \eqref{E_z} and \eqref{sig0 divergent} have the same leading order, while their difference goes as $1/\sqrt{n}$, which converges (slowly). Combining Eqs. \eqref{E_z}, \eqref{sig0} and \eqref{sig0 divergent} then gives the following convergent expression for the image surface charge:
\begin{align}
	\sigma_\text{im}=& \sigma_\text{im}^0 - 2\lim_{z\rightarrow0} \pd_z (V-V^{(0)}) \nonumber\\
	 =& \sigma_\text{im}^0 - \sum_{n=1}^\infty\left(4\frac{J_0(k_n\rho)}{J_1(k_n)^2} - 2\pi^2 J_0((n-1/4)\pi\rho) \right). \label{sig}
\end{align}
which converges except at the rings $\rho=2,4,6,8...$, and is plotted in figure \ref{surface charge}, and compared to the approximation \eqref{sig0}, and to a numerical limit of \eqref{E_z}, that is $2E_z$ for a very small value of $z=0.0001$. All three show a similar pattern - diverging U shaped regions of alternating sign, with relatively smooth regions in between, which become steeper as $\rho$ increases. The singular rings and the source point charge cannot be seen here as they have no thickness along $\rho$.  As noted earlier for the region $0<\rho<2$, $\sigma_\text{im}$ is zero, while $\sigma_\text{im}^0$ is small but finite in this region.  This surface charge is represented schematically in figure \ref{schematic} with blocks of color matching sign.

\section{Charged tight torus}

This problem has a one to one correspondence with the potential of a charged conducting tight torus in free space. This is realized through radial inversion/ Kelvin transform, the transformation $r\rightarrow1/r$, which takes the outside of the cylinder to the inside of the torus. 
The potential transforms as $V(r,\theta,\phi)\rightarrow 1/r~V(1/r,\theta,\phi)$, and by applying this transformation to \eqref{besselsum} gives
\begin{figure}[h!]
	\includegraphics[scale=.67]{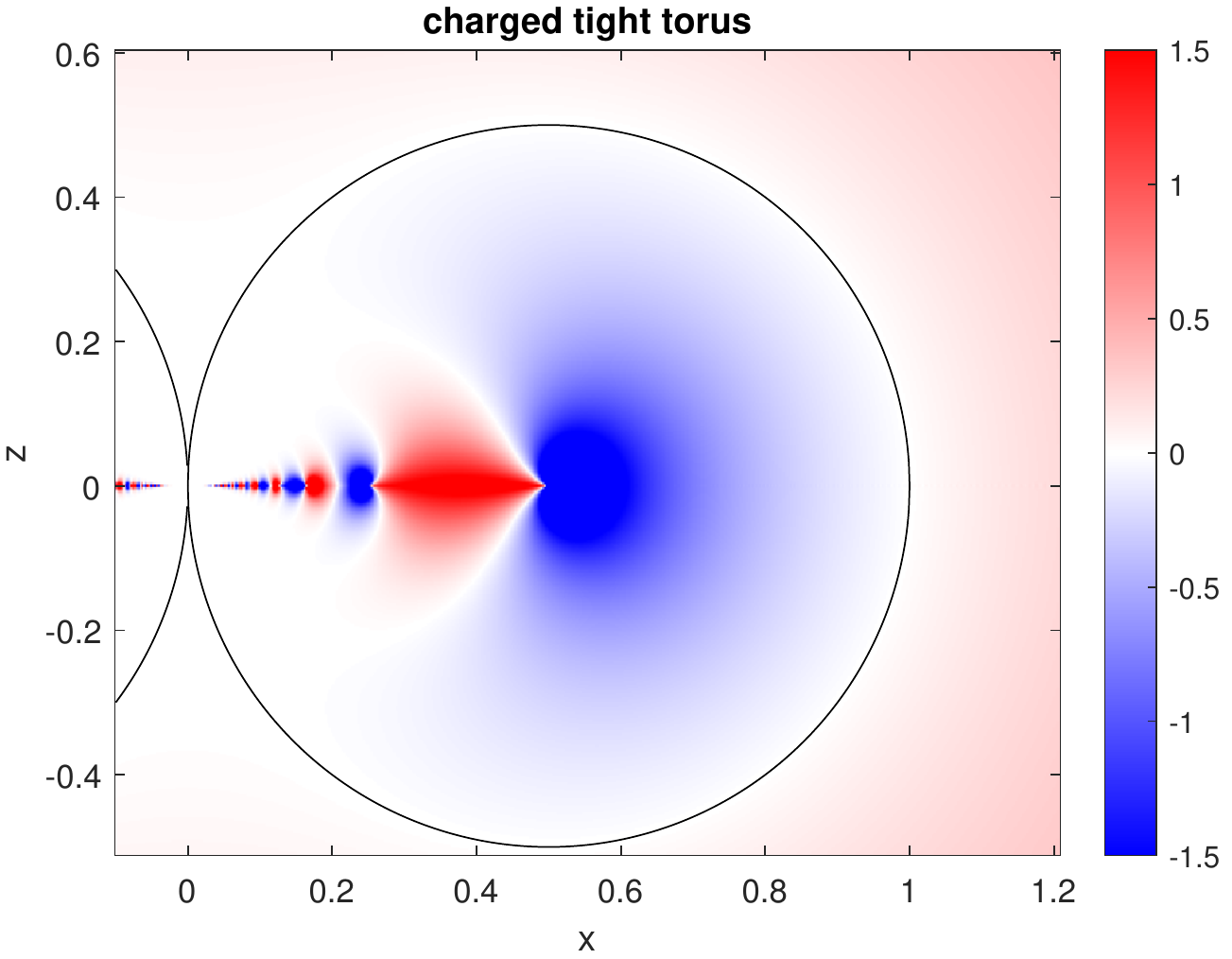}
	\caption{Analytic potential of a charged conducting tight torus as computed by radial inversion of the series \eqref{besselsum}.}\label{torus}
\end{figure}
\begin{align}
V_\text{torus}=\frac{2}{r}\sum_{n=1}^\infty \frac{J_0(k_n(\rho/r^2) e^{-k_n|z|/r^2}}{J_1(k_n)^2}.
\end{align}

which is plotted in figure \ref{torus}, and reveals a singular disk for $\rho<1/2$ - we will call this singularity the `image' of the torus in free space. The Kelvin transform modifies the location of the image by $\rho\rightarrow1/\rho$, so that the image of the torus lies on the disk $\rho<1/2,~z=0$ with singular rings located at $\rho=1/2,1/4,1/6...$. To calculate the corresponding image surface charge density $\sigma_\text{torus}$, we need the limit of the electric field approaching $z=0$:
\begin{align}
\sigma&_\text{im} = 2\lim_{z\rightarrow0}E_\text{z}=\nonumber\\
&-\sum_{n=1}^{\infty} \frac{4e^{-k_n|z|/r^2}}{r^5k_nJ_1(k_n)^2} 
\bigg(  ( k_n(\rho^2-z^2) + |z|r^2 )J_0(k_n\rho/r^2)\nonumber\\ 
&\hspace{3.6cm}- 2k_n |z|\rho J_1(k_n\rho/r^2)  \bigg), \label{E_z torus}
\end{align}
but as for the cylinder, the series diverges for $z=0$, so we follow the above approach of taking the difference of two diverging series. 
The result is
\begin{align}
	&\sigma_\text{torus}=\sigma_\text{torus}^0 \nonumber\\ 
	&+ \frac{1}{\rho^3}\sum_{n=1}^\infty  4\frac{J_0(k_n/\rho)}{J_1(k_n)^2}  - 2\pi^2 (n\!-\!1/4)J_0((n\!-\!1/4)\pi/\rho),   \label{sig torus}
\end{align}
where
\begin{align}
\sigma_\text{torus}^0= 4\sum_{k=1}^\infty \Re\bigg\{\frac{i^k/(2k)^2}{(\rho^2-1/(2k)^2)^{3/2}} \bigg\}. \label{sig0 torus}
\end{align}
And is plotted in figure \ref{surface charge torus}, showing close agreement with the electric field evaluated at $z=0.0001$, except near the singular rings. This discrepancy has been checked to decrease as $z\rightarrow0$ as it should. The magnitude of $\sigma_\text{torus}$ is far higher than for the cylinder, but the surface area is also much smaller.

 For more analysis on the tight torus, see \cite{belevitch1983torus}. The image for a non-tight torus in free space is not solved \cite{majic2018relationships}, and maybe this solution can provide a starting point.\\

\section{Eccentric point charge}

\begin{figure}[H]
	\includegraphics[scale=.75]{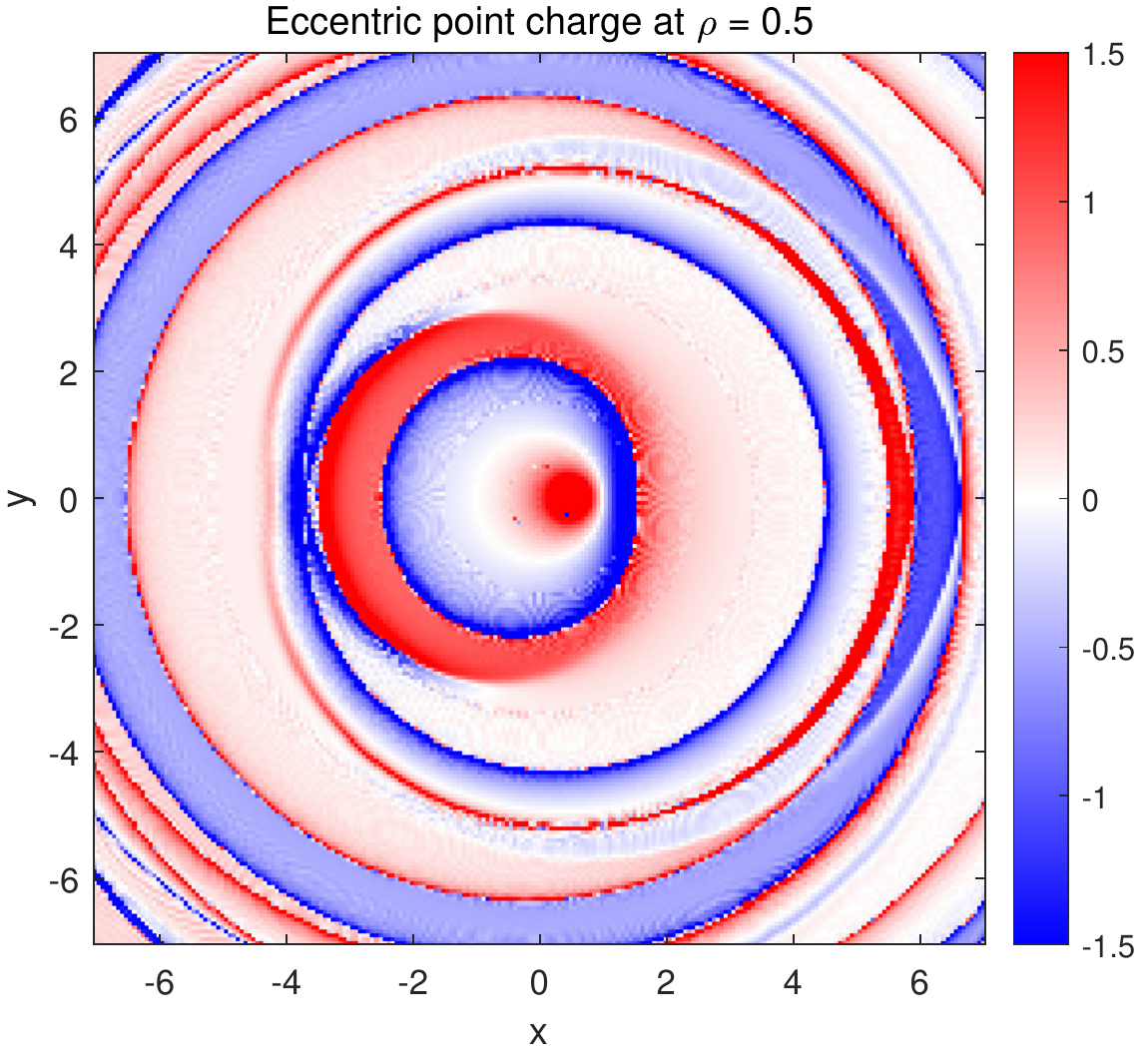}
	\caption{Potential  for an eccentric point charge at $x=0.5,y=0$, in a conducting cylinder of radius 1. $V_\text{eccentric}$ was computed with to $m=90$ and $n=340$ terms, enough to converge to reveal the main features of the image, but with some noise.}\label{eccentricplot}
\end{figure}

For a point charge located at $\rho=\rho_0$ with azimuthal angle $\phi=0$, the solution as a series of Bessel functions is derived in \cite{bouwkamp1947electrostatic,peskoff1974green} to be (correcting a typo in \cite{bouwkamp1947electrostatic} eq. 42):
\begin{align}
	V_\text{eccentric}=&2\sum_{m=0}^\infty\sum_{n=1}^\infty (2-\delta_{m0})\frac{J_m(k_{nm}\rho_0)J_m(k_{nm}\rho)}{k_{nm}J_{m+1}(k_{nm})^2}\nonumber\\
	&\times\cos(m\phi)e^{-k_{nm}|z|}, \label{Veccentric}
\end{align}
where $k_{nm}$ is the $n^{th}$ zero of $J_m$. This formula converges everywhere except on $z=0$ where convergence is non-uniform. $V_\text{eccentric}$ is plotted on the plane $z=0$ in figure \ref{eccentricplot} to reveal the image structure, which is extremely complicated, consisting of distorted image rings which overlap chaotically. 

\section{conclusion}
We have investigated the general form of the image of a point charge at the center of a conducting cylinder, found an exact expression for the singular part of the image, derived a simple and accurate approximation which matches the inner part of the image, and found an exact analytic formula for the image surface charge. The analysis in this problem should hopefully provide insight into the image solutions for similar electrostatic boundary problems - for example toroids in free space or excited by a point source, and prolate or oblate spheroids with a point charge placed in the center.

\bibliography{../libraryH} 

\end{document}